# Automatic labeling of molecular biomarkers of whole slide immunohistochemistry images using fully convolutional networks

Fahime Sheikhzadeh [a,b*], Martial Guillaud [b], Rabab K. Ward [a]
[a]University of British Columbia, Dept. of Electrical Engineering, Vancouver, BC, Canada.
[b]British Columbia Cancer Research Centre, Imaging Unit, Dept. of Integrated Oncology, Vancouver, BC, Canada

**Abstract**. This paper addresses the problem of quantifying biomarkers in multi-stained tissues, based on color and spatial information. A deep learning based method that can automatically localize and quantify the cells expressing biomarker(s) in a whole slide image is proposed. The deep learning network is a fully convolutional network (FCN) whose input is the true RGB color image of a tissue and output is a map of the different biomarkers. The FCN relies on a convolutional neural network (CNN) that classifies each cell separately according to the biomarker it expresses. In this study, images of immunohistochemistry (IHC) stained slides were collected and used. More than 4,500 RGB images of cells were manually labeled based on the expressing biomarkers. The labeled cell images were used to train the CNN (obtaining an accuracy of 92% in a test set). The trained CNN is then extended to an FCN that generates a map of all biomarkers in the whole slide image acquired by the scanner (instead of classifying every cell image). To evaluate our method, we manually labeled all nuclei expressing different biomarkers in two whole slide images and used theses as the ground truth. Our proposed method for immunohistochemical analysis compares well with the manual labeling by humans (average F-score of 0.96).

**\*First Author**, E-mail: fahime@ece.ubc.ca

## 1 Introduction

This study addresses the automation of Immunohistochemistry (IHC) image analysis in quantitative pathology. IHC staining is a widely used tissue based biomarker imaging technology. A tissue slide is stained with different IHC biomarkers. The whole slide is then imaged by a scanner or a CCD color camera mounted on an optical microscope. IHC imaging is widely used in the detection of dysplastic cells and indicates the physiological state and changes in the disease process. It can determine the spatial distribution of every biomarker with respect to the tissue histological structures as well as the spatial relationships amongst the cells expressing each biomarker. This information could provide valuable insight into the molecular complexity of the disease process, enable new cancer screening and diagnosis tools, inform the design of new treatments, monitor treatment effectiveness, and predict the patient response to a treatment. However, the manual quantification and localization of the different subset of cells containing a biomarker, in each IHC stained tissue, demands a large amount of time and effort. In addition, qualitative or semi-quantitative scoring systems suffer from low reproducibility, accuracy, sensitivity and inability to distinguish the different stains apart when they overlap spatially. Therefore, there is a high need for methods that can automatically quantify IHC images. This is however a challenging problem due to the variations caused by the different staining and tissue processing procedures.

The techniques proposed in the literature for quantifying histopathology images are generally based on traditional image processing approaches that rely on extracting pre-defined features of nuclei or cells appearance[1,2]. Using these pre-defined features, in IHC image analysis, raises the concern of low reproducibility and lack of robustness resulting from the variations in nuclei and cell appearances caused by the different staining and tissue processing procedures used in different laboratories. Recently, several studies have shown that deep learning algorithms such as those employing Convolutional Neural Networks (CNN)[3], could be used in histopathology image analysis with great success[4–10]. For instance, Jain et al.[4] have applied CNN to segment electron microscopic images into intracellular and intercellular spaces. Ciresan et al.[5] and Wang et al.[9] employed deep learning to respectively detect mitoses metastastic breast cancer in histopathology images of breast. This approach has outperformed other methods by a significant margin[5,9]. Janowczyk et al.[10] demonstrated that the performance of their CNN approach in segmentation (nuclei, epithelium, tubule and invasive ductal carcinoma), detection (lymphocyte and mitosis) and classification (lymphoma subtype) applications, was comparable or superior to existing approaches or human performance.

In this paper, we propose a deep learning based method that can detect and locate cells that express different biomarkers present in a whole slide image (WSI). The proposed deep learning network is an extension of a convolutional neural network that is trained (using RGB images of cells) to classify a cell in a slide according to its label (expressed biomarker). The deep learning network is a fully convolutional neural network (FCN) based on the trained CNN. The FCN detects and locates the different biomarkers presented in all cells of a WSI.

The benefit of using the deep learning approach (compared to image processing methods that rely on capturing pre-defined features of nuclei appearance) is that the feature descriptors are automatically learned in the convolution kernels[11] (when training the network). Therefore, there is no need to find and quantify sophisticated features for every group of cells expressing a specific biomarker. We believe that by eliminating the need to find and quantify advanced features for each cell expressing a biomarker, we could develop more robust, accurate, and reproducible methods for biomarker localization in IHC images. Also, in our proposed approach, except for a simple averaging normalization, there is no need for pre-processing the RGB images before feeding them to the CNN. This is in contrast to most methods in the literature as these methods rely on complicated processing like un-mixing techniques (e.g. un-mixing multi-spectral images and color un-mixing in RGB images).

In this paper, first we explain how we have collected our dataset which is composed of whole slide images of cervical biopsies. Then we discuss the methods we have used to detect and localize the different biomarkers in WSIs. After that we explain the experimental design, results and validation metrics we used to evaluate our proposed approach.

## 2 Dataset

*2.1. Specimens*

We obtained cervical biopsy specimens from baseline patient visits for colposcopy review at the Women's Clinic at Vancouver General Hospital as described in Sheikhzadeh et al.[12]. Patients would have been referred to the clinic based on a prior abnormal Pap smear test result. All cases were diagnosed as having low grade Cervical Intraepithelial Neoplasia (CIN), high grade CIN, or no abnormality (normal cervix). Uterine cervix biopsies were collected by a trained gynaecologist from clinically selected sites. Biopsy specimens were immediately placed in saline and transferred to the hospital pathology lab in formalin fixative. Each biopsy sample was embedded in paraffin and nine 4-µm transverse sections were cut and mounted on slides for each case[12].

*2.2 Staining and imaging the specimens*

Out of the nine slides cut from each biopsy, slides 1, 5, and 9 had been stained with Hematoxylin and Eosin (H&E) and reviewed by an expert pathologist (DVN) to establish disease grade (including normal, reactive atypia, CIN1, CIN2, and CIN3). Slide 3 was double immunostained for p16/ki-67, using the CINtec PLUS kit (Roche mtm laboratories, Mannheim, Germany) according to the manufacturer's instruction and then counterstained with Hematoxylin. This slide was used for immunohistochemical analysis. To ensure that our analysis is not biased by one staining or tissue processing procedure, each specimen was randomly assigned to one of two pathology technicians. The technicians were asked to perform the staining on different days.

Each of the immunostained tissue samples in our study contained three labels: 1) the counterstain that labels every nucleus (Hematoxylin [H]), 2) the chromogen (fast red) that labels a nucleus bound IHC biomarker (the Ki-67 antigen), and 3) the chromogen (DAB) that visualizes the other IHC biomarker (the p16 antigen), which is localized in the nuclei and cytoplasm[13]. The Ki-67 is one of the most commonly used biomarkers of cell proliferation. It is routinely applied to biopsies of cancerous or pre-cancerous cervical tissues[14]. The p16 protein is an important tumor suppressor protein that regulates the cell cycle. It is also commonly used as a diagnostic biomarker of cervical neoplasia[15].

After staining, we scanned the obtained IHC slides by Pannoramic MIDI scanner (3DHISTECH Ltd., Budapest, Hungary). Using Pannoramic Viewer (3DHISTECH Ltd.) we located the regions exhibiting the worst diagnosis on the H&E slide, and marked the same diagnostic areas on the corresponding CINtec stained slide. Then, the marked areas were exported to RGB images (TIFF format) with a resolution of 1024 by 1024 pixels. These images were used in our analysis and validation. For simplicity, we will refer to these images as whole slide images (WSIs) in the rest of this paper.

## 3 Biomarker labeling method

Figure 1 outlines our proposed approach, which is done in two main steps: i) training a CNN for labeling the expressed biomarkers in one cell, and ii) transforming the trained CNN to FCN to locate all cells expressing each of the different biomarkers in a WSI.

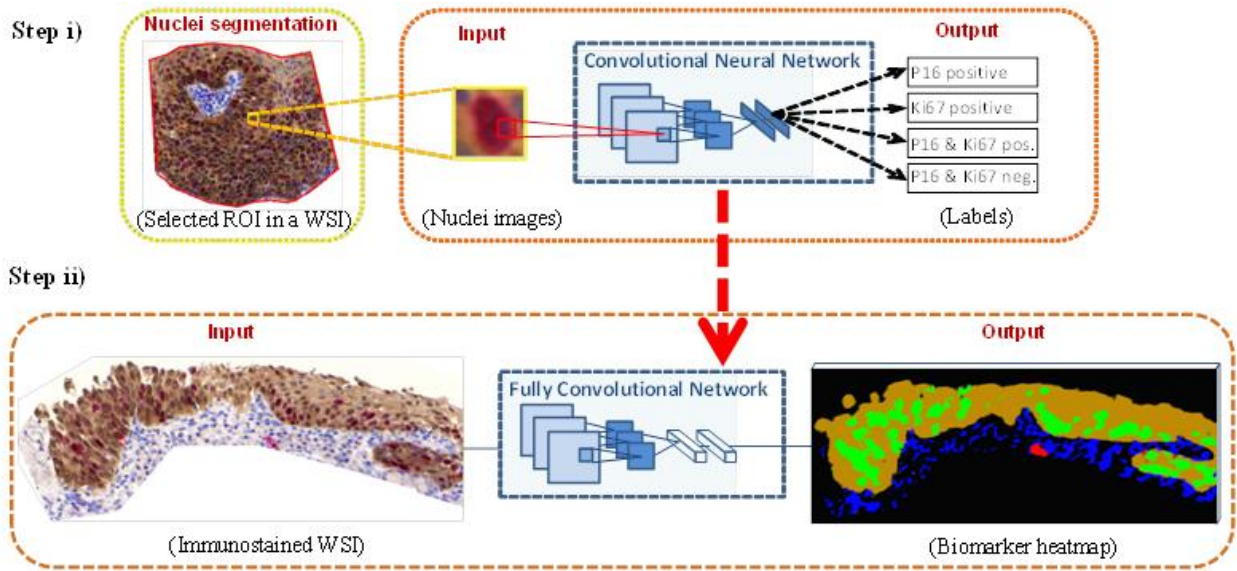

**Fig. 1.** Top: Training a convolutional neural network with nuclei images expressing different biomarkers. Each image has one nucleus only. These nuclei images are segmented from the WSIs in our training set. The CNN learns to classify each nucleus image based on its expressed biomarker(s). Bottom: extending the trained classifier (CNN) to a fully convolutional network. This end to end, pixel to pixel network localizes biomarkers of the whole slide images. The input to this network is a WSI and the output is a heat-map of the different biomarker expression.

In the first step, the nuclei in the IHC images were segmented and used as inputs to the CNN (see Sec. 4.1.1). The output of the CNN was the label that determined the expressed biomarker in the nucleus of each input image. In the second step, a whole slide image of arbitrary size was fed into the network, and for each pixel, the probability of each biomarker being expressed was calculated. At the end, these probability values were used to produce a color heat-map of the different regions expressing different biomarkers in the WSI (see Fig. 1. bottom).

*3.1 Nuclei classification using convolutional neural network*

To build a classifier for labeling the nuclei images, we employed convolutional neural networks[3]. The CNN used in this study consisted of five layers (see Fig. 2). The input to the first layer is an RGB image containing one nucleus. The first layer is a convolution layer which convolves the input image with K filters each having a kernel size of F1 by F1. These kernel matrices are learned by the CNN. The second layer is a max-pooling layer with a kernel size of F2 by F2 which is a sub-sampling layer that reduces the size of the image. The third and forth layers are fully connected layers, each containing M neurons. The output layer consisted of N neurons each with a Gaussian function. This layer generates the probability labels. Figure 2 shows the schematic representation of this CNN.

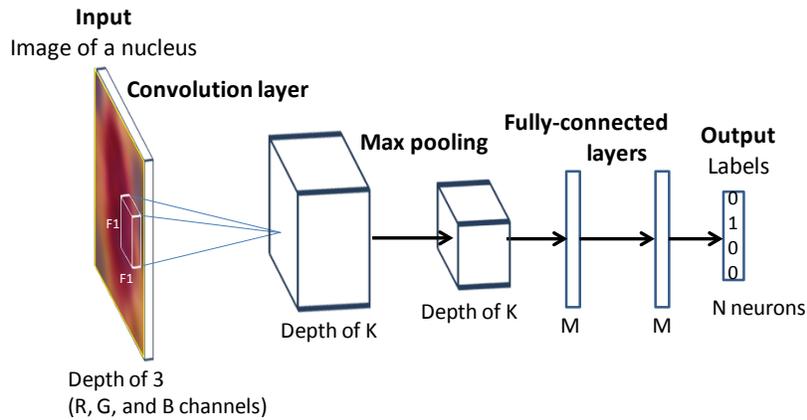

**Fig. 2.** Schematic representation of the architecture of our convolutional neural network. The CNN consists of one convolution layer, followed by a max-pooling layer and two fully connected layers. The last layer is the output layer. The CNN takes an RGB image of a nucleus as input and generates a label as the output.

*3.2 Biomarker localization in WSIs using fully connected network*

In the previous section, we explained how the CNN was trained as a classifier for images, each containing one nucleus (see Sec. 3.1). In order to employ this classifier for localizing biomarkers on a whole slide image (that contains thousand of nuclei), different approaches could be used. For example, a sliding window (whose size is equal to the CNN input size) could be employed on each pixel of the WSI. Taking the window region as the input of the CNN and obtaining the CNN output, could generate a label for each pixel in the WSI. In this study, we took advantage of the nature of convolution which could be considered as a sliding window[16]. By transforming the trained CNN to a fully convolutional neural network, a classification map of the whole slide image is obtained (rather than generating a single classification of a nucleus image by the CNN). This transformation is shown in Fig. 3. As in reference [16], we take the fully connected layers of the CNN and convert them to convolutional layers. This is for the purpose of modifying these layers to make them independent of the spatial size.

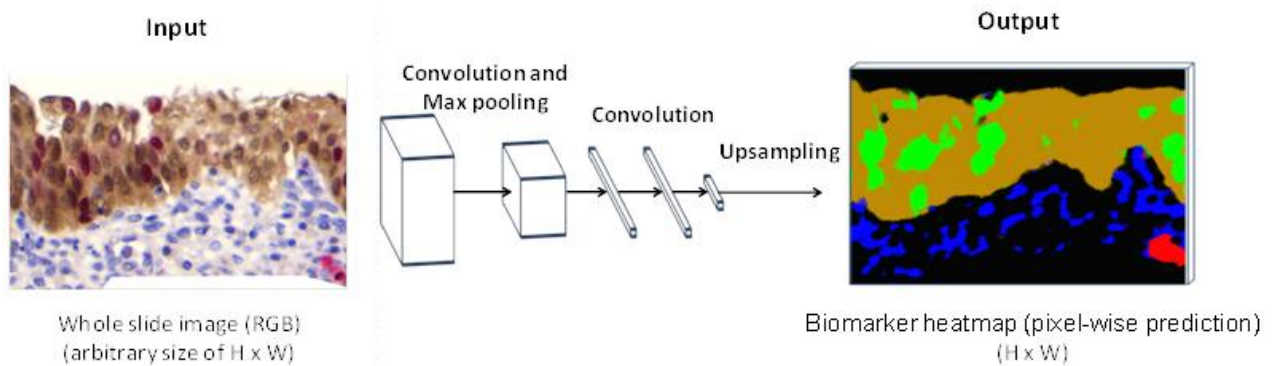

**Fig. 3.** Transforming the fully connected layers in the CNN to convolution layers. This enables the fully convolutional network to produce a pixel-wise output for prediction.

## 4 Experiments and results

*4.1 Biomarker localization*

*4.1.1. Preparing nuclei images for training the CNN*

In order to prepare nuclei images for training the CNN, we first selected different regions of interest (ROI) in six WSIs and then coarsely segmented the nuclei in each ROI. Figure 4 shows the ROI selection, the nuclei images and the label corresponding to each of the different protein expressions.

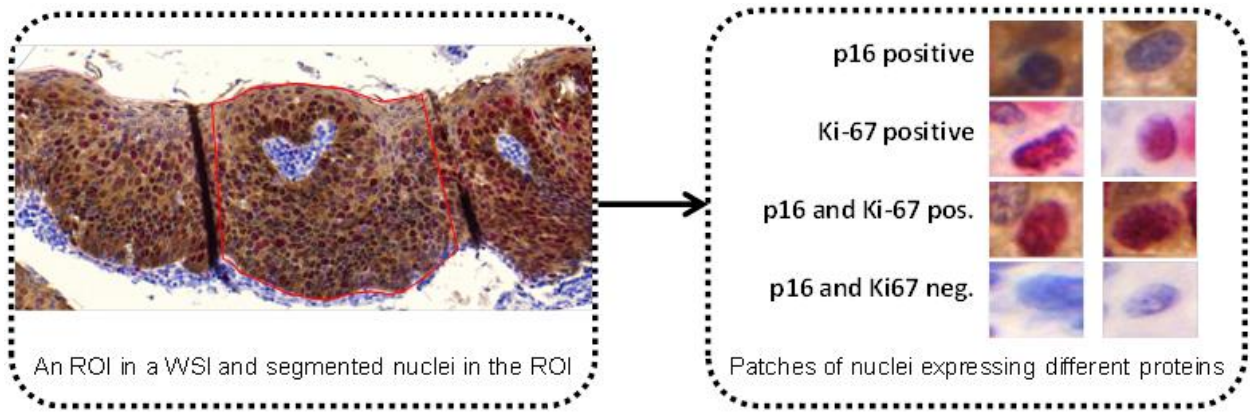

**Fig. 4.** Left: example of an ROI selected in a WSI, within which we segment the nuclei to obtain nuclei images for training the CNN. Right: nuclei images expressing different proteins.

To select the ROIs we used our in-house image analysis software, Getafics[17], and manually delineated the basal membrane and the superficial membrane within each image file. Then, we detected the cell nuclei and their centers of gravity within each ROI at magnification 20X. This procedure was semi-automated, requiring only minor manual adjustments (described previously[18]).

Digital RGB images, each containing one nucleus, were rescaled to 32 by 32 pixel images and stored in a gallery. We then manually labeled the nuclei into four groups: i) positive for p16, ii) positive for Ki-67, iii) positive for both Ki-67 and p16, and iv) negative for both p16 and Ki-67. These nuclei images along with their group number were considered as the ground truth and used for training the CNN.

In total, we had 4679 labeled nuclei RGB images obtained from 6 IHC images. The IHC images were acquired from 6 cases: two CIN3, one CIN2, two CIN1 and one negative cases. From these nuclei, 156 nuclei images obtained from a WSI were used for testing the performance of the CNN, and the remaining 4523 nuclei images were used for training. Table 1 illustrates the distribution of the nuclei in the training and test sets of the different groups.

**Table 1.** Distribution of nuclei in the training and test sets.

|  | Number of nuclei in the different groups | | | |
|---|---|---|---|---|
|  | **p16 positive** | **Ki-67 positive** | **Ki-67 & p16 positive** | **p16 and Ki-67 negative** |
| **Train Set** | 918 | 866 | 1572 | 1167 |
| **Test Set** | 62 | 25 | 52 | 17 |

*4.1.2 Training and testing the CNN*

We used the following network configuration. As mentioned above, the first layer was a convolutional layer containing 96 filters with a kernel size of 11 by 11 and stride of 4. The second layer was a max-pooling layer with a kernel size of 3 by 3 and stride of 2. The third and forth layers were fully connected layers, each containing 2048 neurons. The output layer consisted of four neurons. The CNN in this study was implemented by using the open source deep learning framework Caffe[19], and a system with an Intel I5-6600 3.3GHZ Quad core CPU and a GTX750TI GPU was used.

For training, we used the back-propagation algorithm whereby these labeled nuclei images were the inputs to the CNN. Softmax function is used in the output layer of the CNN. We considered four labels (representing the four groups) as the output of the CNN. The trained CNN learns the features corresponding to the presence of following biomarkers in each nucleus: i) H and p16 (label: 0 0 0 1) ii) H and Ki-67 (label: 0 0 1 0) iii) H and both Ki-67 and p16 (label: 0 1 0 0), iv) H only (absence of Ki-67 and p16) (label: 1 0 0 0).

For training, we stopped after 8000 iterations (with an error rate of 0.04 on the training set). Then we performed the classification on the test set, using the trained CNN. Table 2 shows the confusion matrix of the results of the test set. According to this confusion matrix, the classification rate for the trained CNN was 92% on the test set (error = 0.07).

**Table 2.** Confusion matrix of the classification results on the test set using CNN.

|  | p16 pos. (Predicted) | Ki-67 pos. (Predicted) | Ki-67 & p16 pos. (Predicted) | p16 & Ki-67 neg. (Predicted) | Accuracy |
|---|---|---|---|---|---|
| **p16 pos. (Actual)** | 59 | 0 | 1 | 2 | 95.1% |
| **Ki-67 pos. (Actual)** | 0 | 25 | 0 | 0 | 100% |
| **Ki-67 & p16 pos. (Actual)** | 8 | 0 | 44 | 0 | 84.6% |
| **p16 & Ki-67 neg. (Actual)** | 0 | 0 | 0 | 17 | 100% |
| **Overall Accuracy** |  |  |  |  | 92.9% |

*4.1.3 Labeling biomarkers in WSI using FCN*

After training the CNN to label the biomarker expressed in each cell image, we transformed the CNN into a fully convolutional network (FCN) for analyzing the whole slide image. The FCN produces a classification map that covers an arbitrary input size image, not just a single classification. As described in Sec. 3.2 we transformed the trained CNN to FCN based on the method presented by Shelhamer et al in[16]. Next, we fed each WSI as an input to the FCN and obtained a heat-map for each biomarker. In our dataset, this means four heat-maps each illustrating the regions in a WSI that are : i) positive in p16, ii) positive in Ki67, iii) positive in both p16 and Ki-67, and iv) negative in both p16 and Ki67 (only expressing H). Figure 5 shows a WSI and its corresponding heatmaps obtained as the output of FCN by feeding a WSI as the input.

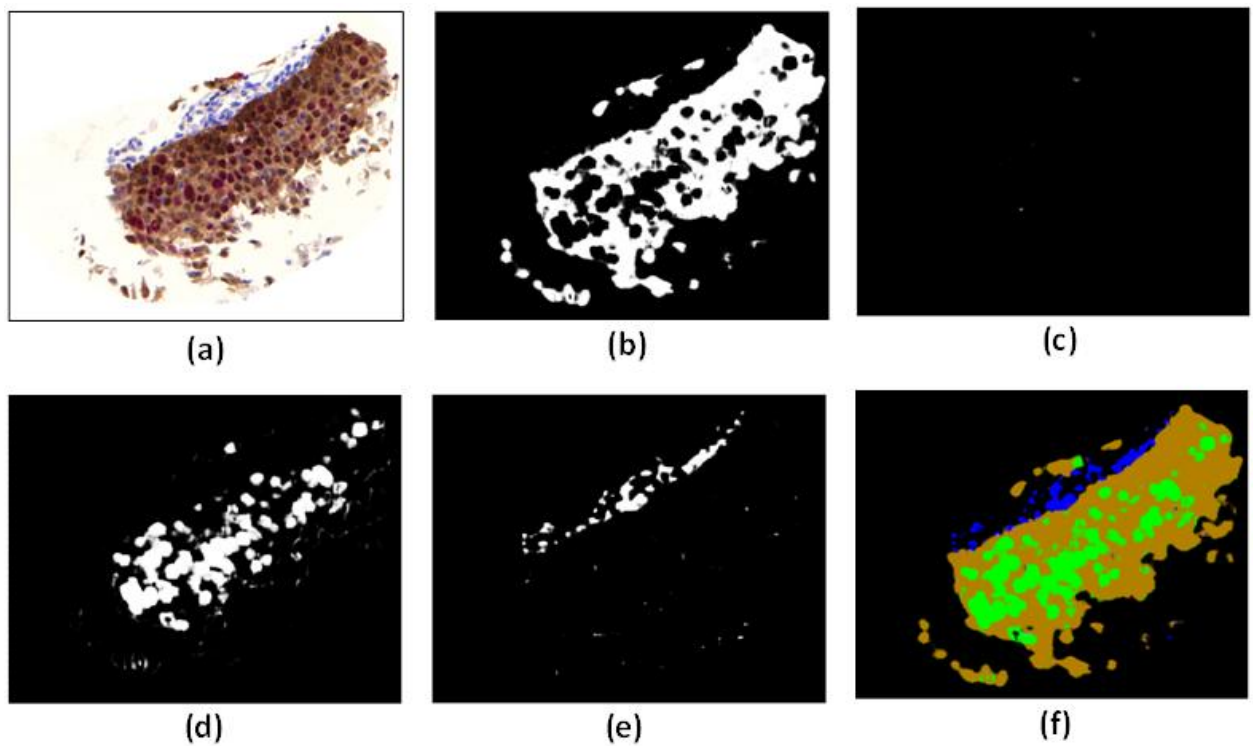

**Fig. 5.** The heat-maps resulted from biomarker labeling on an example WSI using FCN. The WSI of size 2048×2048 obtained from the scanner was fed to the FCN (a). The output of the FCN is the heat-maps (of size 2048 × 2048) marking regions positive in p16 (b), positive in Ki67 (c), positive in both p16 and Ki67 (d), and negative in both p16 and Ki67 (e). By combining these four heat-maps we get the final biomarker heat-map (f).

*4.2 Validation*

To evaluate our proposed biomarker labeling method, we manually labeled the nuclei expressing different biomarkers in WSIs and used them as the ground truth. It is not easy and extremely laborious to manually obtain a biomarker heat-map for each WSI; since this would require manual delineating the membrane of nuclei and labeling each pixel in a WSI. In addition, the knowledge of the exact border of a nucleus or a cell in quantification of an immunostain is not critical and the main concern is to detect the presence of a biomarker in a cell. Hence, we chose to mark only the center of each nucleus and assign a label to the center based on the expressing biomarker(s). To reduce the effects of human error, each WSI was reviewed consequently by three different human readers. Considering the labeled center of a nucleus as the ground truth and comparing it to the biomarker heat-map generated by our proposed method, we defined the following metrics for each biomarker:

- Number of correctly labeled nuclei (true positive): this is the number of nuclei centers in the ground truth which are marked positive for that biomarker and are located inside the biomarker heat-map.

- Number of missed nuclei (false negative): this is the number of nuclei centers in the ground truth which are marked positive for that biomarker and are located outside the biomarker heat-map.

- Number of falsely labeled nuclei (false positive): this is the number of nuclei centers in the ground truth which are not marked positive for that biomarker but are located inside the biomarker heatmap.

From the above metrics, we calculate the precision, recall, and the F-score, which is the harmonic mean of these two[20], described as:

$$Precision = \frac{Truepositive}{Truepositive + Falsepositive}$$

$$Recall = \frac{Truepositive}{Truepositive + Falsenegative}$$

$$F = 2 \times \frac{Precision \times Recall}{Precision + Recall}$$

We selected two WSIs and using our proposed method, we generated their biomarker heat-maps. These images were completely unseen to our network; i.e. they were not used (neither as cell images or whole images) for training or to test the CNN or FCN. Table 3 shows the validation metrics for the FCN performance compared to the human performance (the ground truth).

**Table 3.** FCN performance evaluation of test WSIs: validation metrics of biomarker heat-maps generated by FCN by considering the ground truth as the labels obtained manually by humans.

|  | WSI 1 (high grade CIN) | | | WSI 2 (normal tissue) | | |
|---|---|---|---|---|---|---|
| Image size | 2048×2048 pixels | | | 3072×3072 pixels | | |
| Process time | 166 sec | | | 328 sec | | |
| Biomarker | p16 positive | Ki67 positive | p16 & Ki-67 neg. | p16 positive | Ki67 positive | p16 & Ki-67 neg. |
| True positive | 271 | 157 | 0 | 0 | 18 | 253 |
| False positive | 1 | 12 | 0 | 0 | 2 | 28 |
| False negative | 0 | 0 | 0 | 0 | 0 | 2 |
| Recall | 0.996 | 0.929 | - | - | 0.900 | 0.900 |
| Precision | 1 | 1 | - | - | 1 | 0.992 |
| F-score | 0.998 | 0.963 | - | - | 0.947 | 0.943 |

## 5  Discussion

Immunohistochemical analysis is commonly employed for detecting dysplastic cells in pre-neoplastic lesions and for measuring the spatial distribution of the cells. In IHC images, it is important to localize the cells that express each of the different labels and visualize the different biomarkers. We proposed a deep learning based algorithm for the immunohistochemistry analysis of whole IHC images of an arbitrary size, acquired by whole slide scanners. The results from our method are in accordance with the human observers' classification and compare well with the labeling performed manually by humans.

In this study, we use the color and spatial information of each pixel in a WSI (which is an RGB image) to identify the different biomarkers and localize the cells expressing each of them in one integrated step. In our approach, the color information is not disregarded in training the CNN as color plays an important role in immunohistochemical analysis. Except for a simple averaging normalization, there is no need to perform any pre-processing on the color RGB images. We feed the color WSIs directly to the network. In contrast, most methods in the literature include complicated processing like an un-mixing technique (e.g. un-mixing in multi-spectral images and color un-mixing in RGB images). For example, in the study by Chen et al[6] (in which convolutional neural networks was employed in immunohistochemical analysis), a sparse color unmixing algorithm to unmix the RGB image into different channels representing different markers was used. Then, a CNN was used as a cell detection tool in one channel.

We realize that the decision of labeling cells by a human observer as positive or negative for an immunostain is very subjective, especially in more complex samples such as P16 and Ki67 positive cells. Therefore, we are currently working on improving the methods for validating the deep learning performance in immunohistochemical analysis. For the purpose of this paper visual evaluation of the classification results and comparing them to manually performed labeling were the only available reference.

# 6 Conclusion

We trained an end-to-end network for pixel-wise prediction of molecular biomarkers in immunostained whole slide images. Our results demonstrate that our proposed automatic labeling method compare well (average F-score of 0.96) with the labeling performed manually by humans. The advantage of this method is that it eliminates the need for finding and quantifying sophisticated features of those cells that express a specific biomarker. Therefore, we believe that our approach is more robust and reproducible compared to the methods that rely on extracting pre-defined or hand-crafted features of nuclei or cells appearance. It should be noted that this approach could be easily applied for different tissue types or immunostains with applications in different cancer studies.

*Disclosures*

The authors declare that they have no financial interests or no other potential conflicts of interest in the manuscript.

*Acknowledgments*

We would like to thank Dr. Dirk van Niekerk for reviewing the specimens. In addition, we sincerely thank Ms. Anita Carraro and Ms. Jagoda Korbelic from BC cancer research center for immunostaining and scanning the specimens. This work was funded through a Natural Sciences and Engineering Research Council of Canada grant (RGPIN 2014-04462) and a National Institute of Health funded program project grant (P01–CA-82710).

**Caption List**

**Fig. 1** Our proposed approach for quantifying biomarkers in immunohistochemical analysis.

**Fig. 2** Schematic representation of the architecture of our convolutional neural network (CNN).

**Fig. 3** Our fully convolutional network (FCN) based on the trained CNN.

**Fig. 4** Preparing nuclei images for training the CNN.

**Fig. 5** Results from biomarker labeling on an example whole slide image using our approach.

**Table 1** Distribution of nuclei in train and test sets used for training the CNN.

**Table 2** Confusion matrix of the classification results on a set of nuclei images using CNN.

**Table 3** Evaluation of the performance of FCN for biomarker labeling in test whole slide images.